\documentclass[aps,prb,twocolumn,superscriptaddress,amsmath,nofootinbib,10pt,floatfix]{revtex4-2}
\usepackage{graphicx}
\usepackage[normalem]{ulem}
\usepackage{xcolor}
\usepackage{textgreek}
\usepackage[colorlinks, linkcolor=blue, citecolor=blue]{hyperref}
\usepackage{gensymb}
\usepackage[compat=1.1.0]{tikz-feynman}
\tikzset{
pattern size/.store in=\mcSize, 
pattern size = 5pt,
pattern thickness/.store in=\mcThickness, 
pattern thickness = 0.3pt,
pattern radius/.store in=\mcRadius, 
pattern radius = 1pt}

\usepackage{amssymb}
\newcommand{\be}{\begin{equation}}
\newcommand{\ee}{\end{equation}}
\newcommand{\beq}{\begin{eqnarray}}
\newcommand{\eeq}{\end{eqnarray}}
\newcommand{\ba}{\[\begin{aligned}}
\newcommand{\ea}{\end{aligned}\]}
\newcommand{\bal}{\begin{aligned}}
\newcommand{\eal}{\end{aligned}}

\renewcommand{\vec}[1]{{\bf #1}}
\renewcommand{\epsilon}{\varepsilon}
\renewcommand{\dag}{\dagger}
\def\nn{\nonumber}

\renewcommand{\vec}[1]{\boldsymbol{#1}}

\def \ve{{\varepsilon}}

\def \k{{\vec{k}}}
\def \q{{\vec{q}}}
\def \K{{\vec{K}}}

\def \Q{\vec{Q}}

\def \vp {\Phi}

\def \S{S_{\tn{eff}}}

\def \tn{\textnormal}

\def \ba{\begin{align*}}
\def \ea{\end{align*}}

\newcounter{indice}

\newcommand*{\XW}[1]{\textcolor{black}{ {#1}}}
\newcommand*{\red}[1]{\textcolor{black}{ {#1}}}
\newcommand*{\new}[1]{\textcolor{black}{  {#1}}}
\begin{document}

\title{Interaction-mitigated Landau damping}

\author{Xuepeng Wang}
\affiliation{Department of Physics, Cornell University, Ithaca, New York 14853, USA.}
\author{Roderich Moessner}
\affiliation{Max-Planck-Institut f\"ur Physik komplexer Systeme, N\"othnitzer Stra\ss e 38, 01187 Dresden, Germany}
\author{Debanjan Chowdhury}
\affiliation{Department of Physics, Cornell University, Ithaca, New York 14853, USA.}
\begin{abstract}
Bosonic collective modes are ubiquitous in metals, but over a wide range of energy and momenta suffer from Landau damping, decaying into the continuum of particle-hole excitations. Here we point out that interactions can suppress this decay, protecting a finite fraction of the total spectral weight associated with the collective mode, e.g. a plasmon. The underlying mechanism is level repulsion between a discrete mode and the continuum. We demonstrate the effect using a number of simplified models of strongly correlated Fermi-liquid metals, including a ``solvable" random flavor model in the large$-N$ limit. We discuss in detail the possibility of observing such an avoided decay for plasmons in (moir\'e) graphene-like systems.
\end{abstract}

\maketitle

{\it Introduction.-} Weakly interacting Fermi-liquid (FL) metals have a qualitatively universal excitation spectrum \cite{AGD}. In the vicinity of the sharply defined Fermi surface,  single-particle excitations correspond to the long-lived renormalized quasiparticles. The two-particle excitations associated with the density fluctuations in a neutral FL include: (i) the particle-hole (ph) continuum, which contains significant spectral weight over a range  of frequencies, $0\leq \omega\lesssim  W(\equiv\tn{bandwidth})$, and momenta, $0\leq q\leq 2k_F$ ($k_F\equiv\tn{Fermi momentum}$), and (ii) a gapless collective (zero-sound) mode associated with  fluctuations of the entire Fermi surface \cite{pines}. In a charged FL where electrons interact via Coulomb interactions, the zero-sound mode renormalizes into the plasmon excitation, which can either be gapless or gapped depending on the interplay of dimensionality and screening. The spectral weight in the ph continuum has a sharp onset across a dispersive threshold, $\omega_\star(\q)$; the collective modes in a FL generically enter the ph continuum at sufficiently large  $\q$ and acquire a finite lifetime, decaying via Landau-damping. 
Finding routes to avoid this (kinematically seemingly inevitable) decay of collective modes into the continuum has important experimental and technological implications. For instance, one of the main challenges in the field of plasmonics is tied to the plasmon decay \cite{basov,Ni2018}.

Here, we demonstrate, by a combination of analytics for solvable models and explicit numerics, that a collective mode in an interacting metal can avoid disappearing into the ph continuum. Instead the continuum partially repels it, placing  a finite amount of its spectral weight outside and thereby {\it partially eliminating Landau damping}.  

We exlicitly demonstrate this effect using examples of well-known materials, such as (moir\'e) graphene. While the parameter regimes required to display level repulsion for these materials in an experimentally resolvable fashion might lie outside current capabilities, it seems eminently possible that there exist closely related materials where the effect can be probed directly in the not-too-distant future. The basic  mechanism for such a kinematically allowed \cite{PhysRevB.73.100404, He4} but avoided decay was identified \cite{ruben18} when it was noted that a quasiparticle (e.g. magnon) could be repelled by `its own' two-particle continuum, with repulsion
between a discrete level and a continuum studied in Ref.~\cite{B_Gaveau_1995}.
The present work extends this to the setting of metals, where an abundance of gapless excitations tied to the metallic Fermi surface leads to a continuum that persists down to $\omega\rightarrow0$, which necessitates a careful analysis of the fate of the collective modes. In addition, the plasmon can exhibit an interaction-derived gap, which distinguishes it from Goldstone modes in magnetic systems.

{\it Model.-}  A simple low-energy theory that illustrates the mechanism of level-repulsion of a gapped bosonic collective mode from the ph continuum associated with a Fermi liquid metal has the following  Matsubara action:
\begin{subequations}\label{action}
\beq
S[c^\dag,c,\vp] &=& S_c + S_\vp + S_{\rm{int}}, \label{toymodel}\\
S_c &=& \sum_{k} c^\dagger_{k}( i\omega_n - \ve_\k ) c^{\phantom\dagger}_{k}, \\
 S_\vp &=& \sum_{q} (\Omega_n^2 +\rho_s \q^2 + \Delta^{2}) |\vp_{q}|^2, \\
S_{\rm{int}} &=& \lambda \sum_{k,q} \vp_{q} c^\dagger_{k+q}c^{\phantom\dagger}_{k}.
\eeq
\end{subequations}
The fields $c^\dagger_k,~c_k$ denote the electronic quasiparticle creation and annihilation operators with dispersion $\ve_\k$. The  bosonic collective mode, $\vp$, has a gap $\Delta$, and stiffness $\rho_s$. It is coupled to a ph excitation via a Yukawa coupling of strength, $\lambda$. We use a shorthand notation, $k\equiv(i\omega_n,\k)$ and $q\equiv(i\Omega_n,\q)$, where $i\omega_n ~(i\Omega_n)$ represent fermionic (bosonic) Matsubara frequencies, respectively. 

{\it Analytical results.-} For analytic tractability, we linearize the  dispersion near the Fermi surface, $\ve_\k\approx v_c|\delta\k|$, where $|\delta\k|$ measures the deviation around the Fermi momentum. Tracing out the $c$-electrons yields an effective theory purely in terms of the collective modes:

\begin{subequations}
\beq\label{plrz_tot}
    \label{plrz}
        \S[\vp] &=& \sum_{q}[\Omega^2-(\rho_s \q^2+\Delta^{2})+ \Pi_c(\q,\Omega)]\vp_q \vp_{-q},\nn \label{plrz2}\\ \\
        \Pi_c(\q,\Omega) &=& \frac{\lambda^2 }{v_c}\bigg[1-\frac{\Omega}{\sqrt{\Omega^2-(v_c \q)^2}}\bigg] \label{pic},
\eeq
\end{subequations}
 We  introduce dimensionless units, $\tilde{\lambda}=\lambda^2/v_c^{3}$, $\widetilde{\Omega}=\Omega/v_c$, $\widetilde{\rho}_s=\rho_s/v_c^2$ and $\widetilde{\Delta}=\Delta/v_c$. We  set the lattice constant $a=1$ for convenience; so that $\q$ denotes a dimensionless momentum. \XW{Henceforth we focus specifically on two-dimensional systems; the approach can be similarly extended to higher dimensions in an analogous fashion.} The renormalized $\vp-$propagator can then be expressed as,
 \beq
 D(\q,\Omega)=\frac{1/v_c^{2}}{\bigg[\widetilde{\Omega}^2-(\tilde{\rho}_s \q^2+\tilde{\Delta}^{2}-\tilde{\lambda})- \frac{\tilde{\lambda}\widetilde{\Omega}}{\sqrt{\widetilde{\Omega}^2-\q^2}}\bigg]}.
 \label{Dqo}
 \eeq
 In Fig.~\ref{toy}(a)-(b), we plot the associated spectral function, $\tn{Im}[D(\q,\Omega)]$, as a function of $q,~\Omega$ for two different values of $\tilde\lambda$. 
 
 \begin{figure}[h!]
\includegraphics[width=84mm,scale=1]{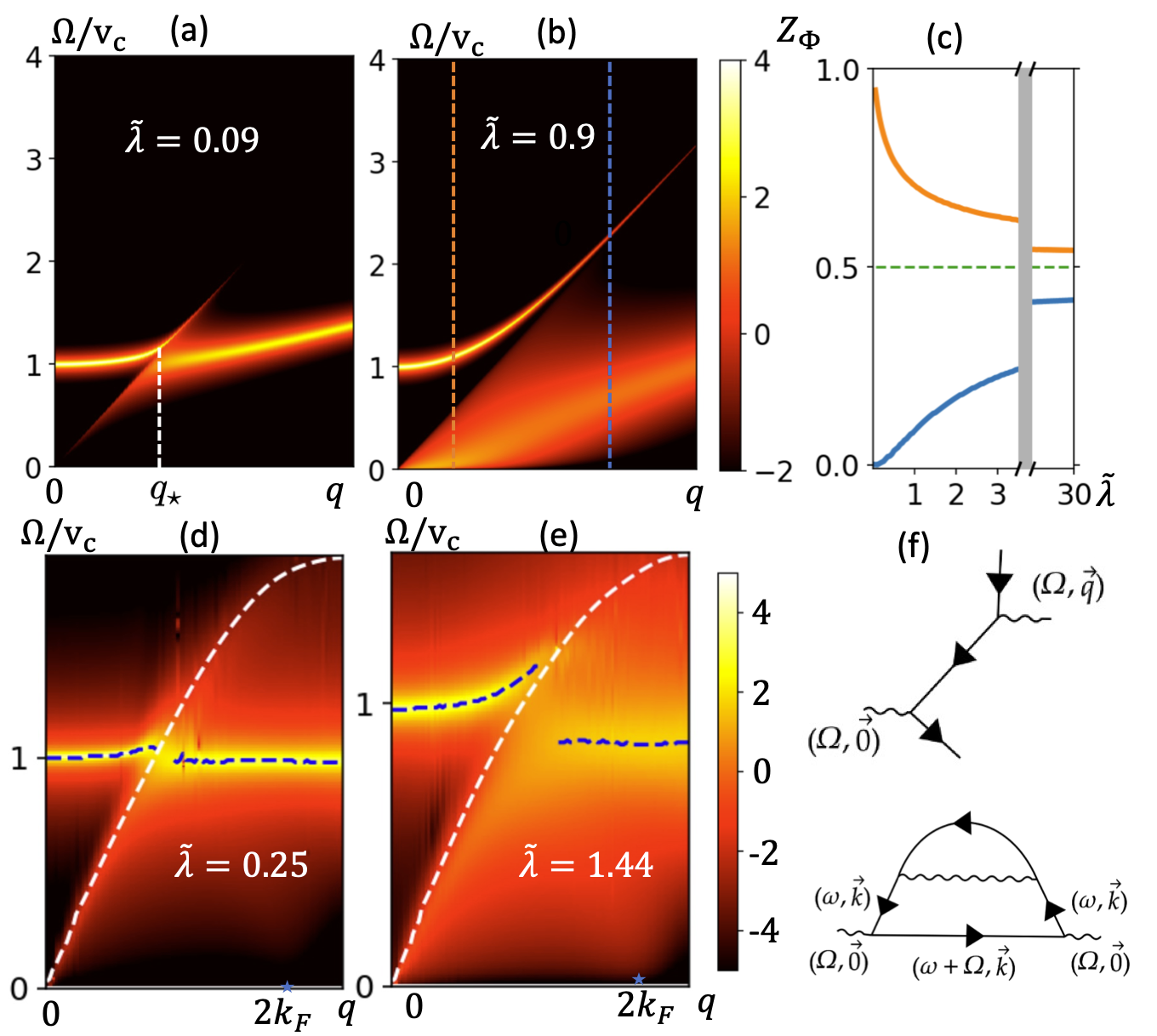}
\caption{\label{toy} 
The response function $\tn{log} [\tn{Im} D(\q,\Omega)]$ in Eq.~\ref{Dqo} obtained from Eq.~\ref{action}, evaluated at (a) weak ($\tilde{\lambda}=0.09$) and (b) strong ($\tilde{\lambda}=0.9$) coupling, respectively. We introduce a small broadening, $\eta=0.01$, for visualization purpose when evaluating $\tn{Im} D(\q,\Omega+i\eta)$. The white vertical dashed line in (a) labels the momentum $q_*$ at which the collective mode crosses the onset of the ph continuum when $\lambda=0$. (c) Numerically obtained $Z_\Phi$ as a function of $\tilde\lambda$ for two different $q$ values (vertical colored dashed lines in (b)). We set $\widetilde\Delta=0.5$ to be in the regime where the asymptotic limit in Eq.\ref{zb_ana} is justified. (d)-(e): Disorder-averaged results for $\tn{log}[\tn{Im} D(\q,\Omega)]$ for the random flavor model with $t_c=1$, $\Delta=1$, $\mu=-1.25 t_c$ and interaction strengths (d) $\lambda=0.5$, and (e) $\lambda=1.2$, respectively. The corresponding $\tilde\lambda$ is defined as described in the main text. White dashed lines denote the onset of the continuum. Blue dashed lines denote the peak associated with $\tn{Im}[D(\q,\Omega)]$, corresponding to the renormalized collective mode. The electronic quasiparticle residue, $Z_{c}$, is obtained self-consistently by solving the saddle point equation \cite{si}, with (d) $Z_{c}\sim 0.9$, and (e) $Z_{c} \sim 0.8$, respectively.  (f) Feynman diagrams for multi-particle scattering processes and the leading order self-energy for the collective mode.}
\end{figure}

 For $\tilde{\lambda}\ll1$, we recover the picture of standard Landau-damping --- the mode remains sharply defined until it enters the continuum and decays by emitting particle-hole pairs, resulting in a broadened dispersive mode (Fig.~\ref{toy}a).  There is a tiny expelled fraction outside the continuum, but the gap between the expelled fraction and the onset of continuum $\sim \tilde\lambda^2/2q^{3}$ \cite{si} which makes it difficult to discern at large $q$. On the other hand, for $\tilde{\lambda}\gg1$, part of the collective mode is repelled outside of the continuum (Fig.~\ref{toy}b), grazing along its edge; the remaining fraction enters the continuum and is damped strongly due to the larger coupling. Thus while it is impossible to avoid decay altogether, an appreciable fraction of the collective mode appears to escape the inevitable damping--this is what we refer to as interaction-mitigated Landau damping.

 Note that the model by construction does not have any information associated with momentum transfer near $2k_F$, as shown in Figs.~\ref{toy}a-b. Next we address two questions, namely (i) what is the maximum spectral weight contained in the undamped branch of the collective mode in the asymptotic limit of $\tilde\lambda\gg 1$, and, (ii) is there a strong-coupling limit where the above picture can be applied using a ``controlled'' computation.    
 We consider the ``quasiparticle-residue'' associated with the collective mode, $Z_\vp(\q)=[1+\partial_{\Omega^{2}} \Pi_c'(\q,\Omega)]^{-1}$, given by
 \beq
 \label{z_eq}
 Z_\vp(\q) &=& \bigg[1+\tilde{\lambda} \frac{\q^2}{2\tilde{\Omega}_{\rm{peak}}(\tilde{\Omega}_{\rm{peak}}^2-\q^2)^{3/2}}\bigg]^{-1},
 \eeq
where $\Omega_{\rm{peak}}$ denotes the renormalized dispersion for the long-lived collective mode {\it outside} the continuum, obtained as the solution to $D^{-1}(\q,\Omega_{\rm{peak}})=0$. The numerical solution \cite{si} for $Z_{\vp}(\q)$ for two different $\q$ values (dashed colored lines in Fig.~\ref{toy}b) is shown in Fig.~\ref{toy}c as a function of $\tilde\lambda$. Clearly, there are qualitative differences in the $\tilde\lambda$ dependence of $Z_\Phi(\q)$ even for the collective mode branch lying outside the ph continuum, depending on whether for the specific $\q$ decay to particle-hole excitations is kinematically allowed.  

Let us denote the crossing-point between the collective mode and the onset of the continuum in the absence of a coupling ($\lambda=0$) as $q_*$ (see Fig.~\ref{toy}a). For the collective mode branch with $q<q_*$, $Z_\vp$ decreases monotonically from $Z_{\vp}(\lambda=0)=1$ with increasing $\tilde\lambda$; see the orange curve in Fig.~\ref{toy}c. On the other hand, for the portion of the branch with $q>q_\star$, we find that $Z_{\vp}$ increases monotonically from $Z_{\vp}(\lambda=0)=0$ with increasing $\tilde\lambda$. After all, the branch does {\it not} exist outside the continuum at infinitesimal coupling, but develops a finite $Z_{\vp}$ once it is pushed outside it. We find that for $\tilde\lambda\gg1$, the spectral weight converges to $Z_{\vp}(\q)\rightarrow1/2$ for both branches, suggesting a universal limit. The analytical expressions in the two  limits is obtained as follows \cite{si}: (i) for $\tilde\lambda \gg 1$,  start from $Z_{\vp}(\q)=1/2$ and include the higher order corrections in powers of $\delta q/\tilde\lambda$, where $\delta q=(2/q^2)|q_\star^2-q^2|^2$, and (ii) for $\tilde\lambda\ll1$ and $q<q_\star$, $Z_{\vp}(\q)$ is computed in the limit $q\ll\widetilde\Delta$, whereas for $q>q_\star$ it is computed in the limit $q\gg\widetilde\Delta$, respectively. Explicitly,
\begin{subequations}
\beq
\label{zb_ana}
Z_{\vp}(q<q_\star) &\simeq& 
\begin{cases}1-\frac{|\q|^2}{2\tilde{\Delta}^{4}}\tilde\lambda, ~~~~~~~\tilde\lambda \ll 1,
\\
\frac{1}{2}+\frac{3}{8}(\frac{\delta q}{\tilde\lambda})^{1/2}, ~~\tilde\lambda \gg 1,
\end{cases}\\
Z_{\vp}(q>q_\star) &\simeq& 
\begin{cases}
\frac{2\tilde\lambda^2}{|\q|^{4}}, ~~~~~~~~~~~~~~\tilde\lambda \ll 1,\\
\frac{1}{2}-\frac{3}{8}(\frac{\delta q}{\tilde\lambda})^{1/2}, ~~\tilde\lambda \gg 1.
\end{cases}
\eeq
\end{subequations}
\red{Note that even when $\lambda\ll1$, a non-zero fraction of the spectral weight is contained in the branch of the collective mode that is repulsed outside the continuum. However, since both  $Z_{\vp}$ and the energy separation between the collective mode and the continuum are relative small ($\sim \tilde\lambda^2$), we focus on the strong coupling case for a clearer illustration of the effect. }

{\it Exact results for random flavor models.-} Given that our results are obtained in an  RPA theory without a small parameter, we now turn to a ``solvable'' model where the analogous computations can be carried out in a controlled setting \cite{DCrmp}.  
We reprise Eq.~\ref{action} and replicate the actions $S_c,~S_{\vp}$ to include $N$ copies of fermions, $c^\dagger_{k,i},~c^{\phantom\dagger}_{k,i}$ ($i=1,..,N$) and $M$ copies of the gapped collective mode, $\vp_{q,\alpha}$ ($\alpha=1,..,M$). We replace the uniform Yukawa-interaction, $S_{\rm{int}}$, by the ``random-flavor'' form \cite{SS22, Fu:2016vas, Patel:2018zpy, Marcus:2018tsr, Wang:2019bpd, JS, Wang:2020dtj, Kim:2020jpz, Adalpe20, WangMeng21}, 
\beq
S_{\rm{int}} \rightarrow \sqrt{\frac{2}{MN}}\sum_{i,i'}\sum_{\alpha}\sum_{r}\lambda_{i,i',\alpha}\vp_{r,\alpha}c^{\dagger}_{r,i}c^{\phantom\dagger}_{r,i'}.
\eeq
Here the couplings $\lambda_{i,i',\alpha}$ are drawn from a random distribution with $\overline{\lambda_{i,i',\alpha}}=0$ and $\overline{\lambda_{i,i',\alpha}^2}=\lambda^2$. We are interested in the large$-N,M$ limit at fixed $N/M$. The RPA equations for the boson and fermion self-energy are exact in this limit for arbitrarily large $\lambda$ \cite{si}. While much of the recent interest in this model is tied to its non-Fermi liquid regime, where the boson is critical, here we focus on the situation where the renormalized boson mass is fixed to be $\Delta^* = v_c$, with $v_c$ the bare Fermi velocity. The metallic system remains in a renormalized Fermi liquid regime at low energies \XW{whenever $\Delta^*$ is finite}. 

Repeating the earlier analysis, we find that a significant spectral weight associated with the bosonic collective mode is repelled outside and continues to graze the edge of the renormalized ph continuum; see Figs.~\ref{toy} d-e. However, with increasing $\lambda$ the mode appears to be broader (i.e. has a larger decay rate). The enhanced broadening is due to multi-particle decay, as shown by the Feynman diagram in Fig.~\ref{toy}f. Note that such processes are automatically included in the self-consistent equations for the boson lifetime, leading to $\Sigma_\Phi''(\Omega)\sim \tilde{\lambda}^{3} \Omega$ at $q=0$ \cite{si}. 

Our computations thus highlight the non-trivial aspects associated with both decay into, and level-repulsion from, a multi-particle continuum in a  renormalized Fermi liquid for arbitrarily strong interactions. While it is rare to find realistic model Hamiltonians where  theoretically well controlled computations can be carried out in the strong-coupling limit, we find that RPA already captures much of the essence of the underlying physics. 

We next turn to  the charge response in graphene, and moir\'e graphene, in the presence of screened Coulomb interactions. Importantly, instead of focusing on an independent bosonic mode, we focus on the intrinsic plasmon excitation and analyze the regime where it is repelled from the continuum.

{\it Charge response in graphene-like models.-} We will begin by analyzing numerically the charge response for the usual honeycomb model of graphene, with the Coulomb interaction treated at the RPA level. The ph continuum in graphene is well known to host an intra and inter-band contribution: $\Pi(\q,\Omega) = 2[\Pi_{\tn{intra}}(\q,\Omega) + \Pi_{\tn{inter}}(\q,\Omega)]$, where the factor of 2 is due to spin degeneracy. The full RPA charge susceptibility is
\beq
\label{graphene_chi}
\chi(\q,\Omega)=\frac{\Pi(\q,\Omega)}{1-V(\q)\Pi(\q,\Omega)}.
\eeq
We  allow ourselves the freedom to tune the strength of Coulomb interaction, $V(\q)$, by effectively varying the ``charge'' of the electron. We will demonstrate that for strong enough Coulomb interaction and heavy doping, the plasmon mode is completely pushed out of the intra-band continuum, enters partially inside the inter-band continuum at a momentum $q_\star$, and is repelled partially away from the inter-band continuum; see Figs.~\ref{graphene_main}a-b. For a 2D Coulomb interaction derived from an underlying one in 3D, $V(\q)=2 \pi e^2/ \kappa q$, where $\kappa$ is an effective dielectric constant. To compare the scale of the Coulomb interaction with the graphene bare bandwidth, $W_0$, we define a dimensionless Coulomb potential \cite{si}, $\widetilde{V}(\q)=2 \pi e^2/(\kappa a_0^2 W_0q) = \tilde{e}^2/ a_0 q$.  The dimensionless interaction strength $\tilde{e}^2 \equiv 2 \pi e^2/ \kappa a_0 W_0$ ($a_0$ is the lattice constant). We are now in a position to obtain an analytical understanding of the level-repulsion in graphene, building on the results of our earlier simple models.

\begin{figure}[h!]
\includegraphics[width=85mm,scale=1]{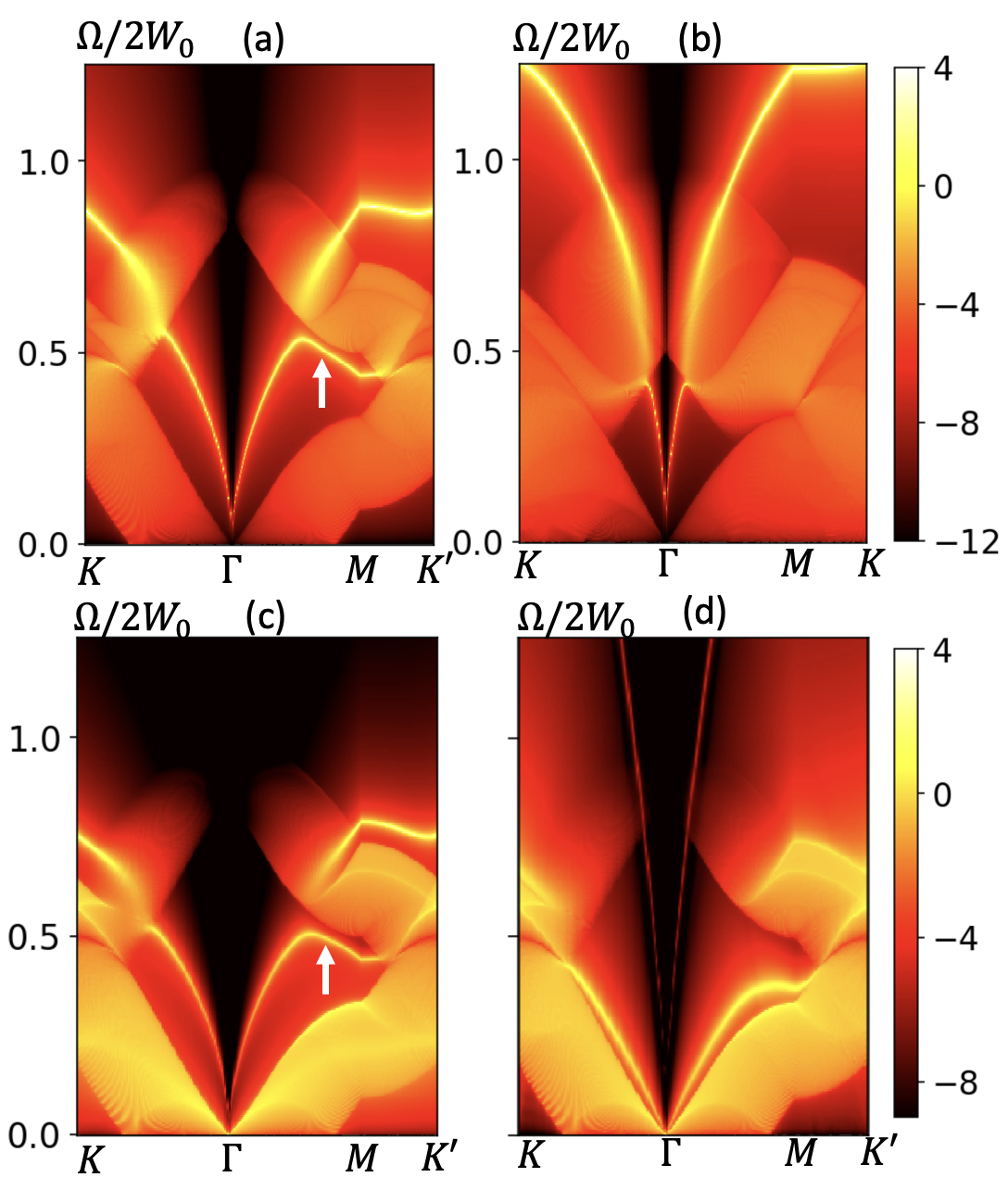}
\caption{\label{graphene_main} Numerical results for the spectral function, $\tn{log}~ \tn{Im}[\chi(q,\Omega)]$, for graphene (without substrate in a, b) with 2D Coulomb interaction with $\tilde{e}^2=15\simeq 6 \tilde{e}^2_{\tn{real}}$. The Fermi-energies are (a) $E_F/W_0 \simeq 0.83$ (b) $E_F/W_0 \simeq 0.5$, respectively, where $E_F$ is measured relative to charge neutrality. The plasmon is repelled from the inter-band continuum, and its dispersion develops a negative slope in the region indicated by the white arrows. Numerical results for $\tn{log}~ \tn{Im}[\chi(q,\Omega)]$ for graphene on a metallic substrate (without direct electron tunneling) coupled through Coulomb interaction with $\tilde{e}^2=7$ for (c) \XW{$v_F^{\tn{sub}}<v_F^{\tn{G}}$, and (d) $v_F^{\tn{sub}}>v_F^{\tn{G}}$}. The red line with the largest slope is the plasmon mode of the substrate.}
\end{figure}

Within RPA and for $q\rightarrow0$ and $(2E_F-\Omega-v_Fq)\rightarrow0$, the asymptotic expressions \cite{pnas,rmp_sarma,sarma2,Wunsch_2006} for $\Pi_{\tn{intra}}(\q,\Omega)$ and $\Pi_{\tn{inter}}(\q,\Omega)$ are given by \cite{si}
\begin{subequations}\label{Pis}  
\beq  
    \Pi_{\tn{intra}}(\q,\Omega)&=&\bigg[1-\frac{\tilde\Omega}{\sqrt{\tilde\Omega^2-|\q|^2}}\bigg],\\   
    \Pi_{\tn{inter}}(\q,\Omega)&\simeq& \pi[(\tilde\Omega-2\tilde E_F)+\sqrt{(\tilde\Omega-2\tilde E_F)^2-|\q|^2}]+\Pi_{\tn{reg}} \nn\\
\eeq
\end{subequations}
where $\tilde\Omega\equiv \Omega/v_F$, $\tilde E_F \equiv E_F/v_F$,and $\Pi_{\rm{reg}}$ includes additional non-singular terms. The pole structure of Eq.~\ref{graphene_chi} in the limit $\Omega\gg |\q|$ is given by,
\beq\label{graphene_asym}
\chi(\q,\Omega)\approx\frac{\Pi(\q,\Omega)}{1-\frac{\beta_0 |\q|}{\tilde\Omega^2}-\frac{\tilde{e}^2}{|\q|}\Pi_{\tn{inter}}(\q,\Omega)}.
\eeq
The $1-(\beta_0 |\q|/\tilde\Omega^2)$ term is obtained upon expanding $V(\q)\Pi_{\tn{intra}}(q,\Omega)$ in powers of $|\q|/\tilde\Omega$, leading to a plasmon mode with dispersion $\Omega\sim\sqrt{\beta_0 |\q|}$  ($\beta_0=\pi e^2/\kappa W_0$), which is coupled to the interband continuum through the Coulomb interaction. Solving for $\tn{Re}\chi^{-1}(\q,\Omega)=0$  yields the dispersion of the mode that avoids repulsion from the inter-band continuum, in a manner analogous to our earlier discussion.

To study the consequences of dynamical screening of the Coulomb interaction on the phenomenology of level-repulsion, we have also analyzed the charge response in a bilayer system where a single sheet of graphene (with a Fermi velocity denoted $v_F^{\tn{G}}$) is coupled via Coulomb interaction to a metallic Fermi liquid substrate. We describe the latter microscopically by a lattice model of electrons on a triangular lattice with Fermi velocity, $v_F^{\tn{sub}}$. We choose the filling such that $E_F^{\tn{sub}}=0.6W_0^{\tn{sub}}$, where $W_0^{\tn{sub}}$ is the electronic bandwidth in the substrate. Note that we exclude  direct hopping of electrons between the two layers, such that there is a $U(1)\times U(1)$ symmetry associated with the two conserved densities. Within RPA, the dynamics of both charge degrees of freedom is coupled in a non-trivial fashion. For $v_F^{\tn{sub}}<v_F^{\tn{G}}$ (Fig.~\ref{graphene_main}c), the phenomenology of level-repulsion is qualitatively similar to where the substrate is absent (Fig.~\ref{graphene_main}a), and the latter only modifies the results quantitatively by screening the Coulomb potential. Importantly, in this case, the onset of the ph continuum for the substrate lies inside the graphene ph continuum. On the other hand, for $v_F^{\tn{sub}}>v_F^{\tn{G}}$ (Fig.~\ref{graphene_main}d), the graphene plasmon decays into the ph continuum of the metallic substrate before entering the graphene interband continuum. As a result the phenomenology changes qualitatively. This illustrates that, in principle, the fate of the graphene plasmon can be determined by tuning the properties (e.g. density, bandwidth) associated with an underlying substrate in a controllable device geometry.

{\it Charge response in twisted bilayer graphene-like models.-} Our results so far suggest that heavily doped graphene ($ E_F/W_0 \sim 0.8$, where $W_0$ is the bandwidth) with larger than usual strength of Coulomb interaction can display an avoided level-repulsion of the plasmon from the inter-band continuum. Let us now turn to the related setup of twisted bilayer graphene, in order to analyze the extent of possible phenomenological similarities. We  focus on angles away from magic-angle, where the isolated bands are {\it not} flat and the effects of Coulomb interactions can be justifiably treated within RPA. 

We begin with the Bistritzer-Macdonald (BM) model \cite{bm, koshinofu, tbg1, tbg3} at twist angle $\theta=1.4$, and include the effect of a screened Coulomb interaction, $V(\q)=(2 \pi e^2/\kappa q) \tanh (\xi q)$, where the background dielectric constant $\kappa\simeq6$ with a screening length $\xi \simeq 10~\tn{nm}$ \cite{si}. The numerical results for the charge response are shown in Fig.~\ref{tbg_1.4}, where we plot the dielectric function \cite{si,local1, local2} to leading order, after projecting to the flat bands,
\beq
\epsilon^{-1}_{\vec{G}}(\q,\Omega)=\bigg[1-\sum_{\eta}V(\q+\vec{G})~\Pi^{\eta}_{\vec{G}}(\q,\Omega)\bigg]^{-1},
\label{df}
\eeq
with $\eta$ denoting the valley index and $\vec{G}$ the reciprocal lattice vector in the moir\'e Brillouin zone. 
Interestingly, the experimentally realistic value of Coulomb interaction does not immediately repel the plasmon from the continuum (Fig.~\ref{tbg_1.4}a), as was the case for single layer graphene. However, if the strength of Coulomb interaction is increased (by a factor of 4),  a clear signature appears of plasmon level-repulsion from the inter-band continuum (Fig.~\ref{tbg_1.4}b) for the density shown in Fig.~\ref{tbg_1.4}c. A fraction of the plasmon is repelled outside the inter-band ph continuum near $q_*$ (marked by white arrows), while the rest enters the continuum (marked by blue arrows) and is Landau damped. For $\Omega\gtrsim70~ \tn{meV}$, the plasmon exits the inter-band ph continuum and remains undamped. Although the plasmon above the inter-band ph continuum may finally decay into the continuum associated with the remote bands, the fraction of plasmon repelled below the inter-band continuum remains undamped \cite{si}. While the latter mechanism was studied in previous work \cite{pnas}, our observations point out a distinct mechanism to avoid plasmon decay via level-repulsion from the continuum.
\XW{As long as the interaction between the plasmon and the inter-band continuum is kinematically allowed (e.g. achieved by increasing the strength of the effective Coulomb interaction), the mechanism of level-repulsion discussed here can appear in tandem with the unrelated mechanism pointed out in previous work \cite{pnas} at different energy scales. }

\begin{figure}[h!]
\includegraphics[width=90mm,scale=1]{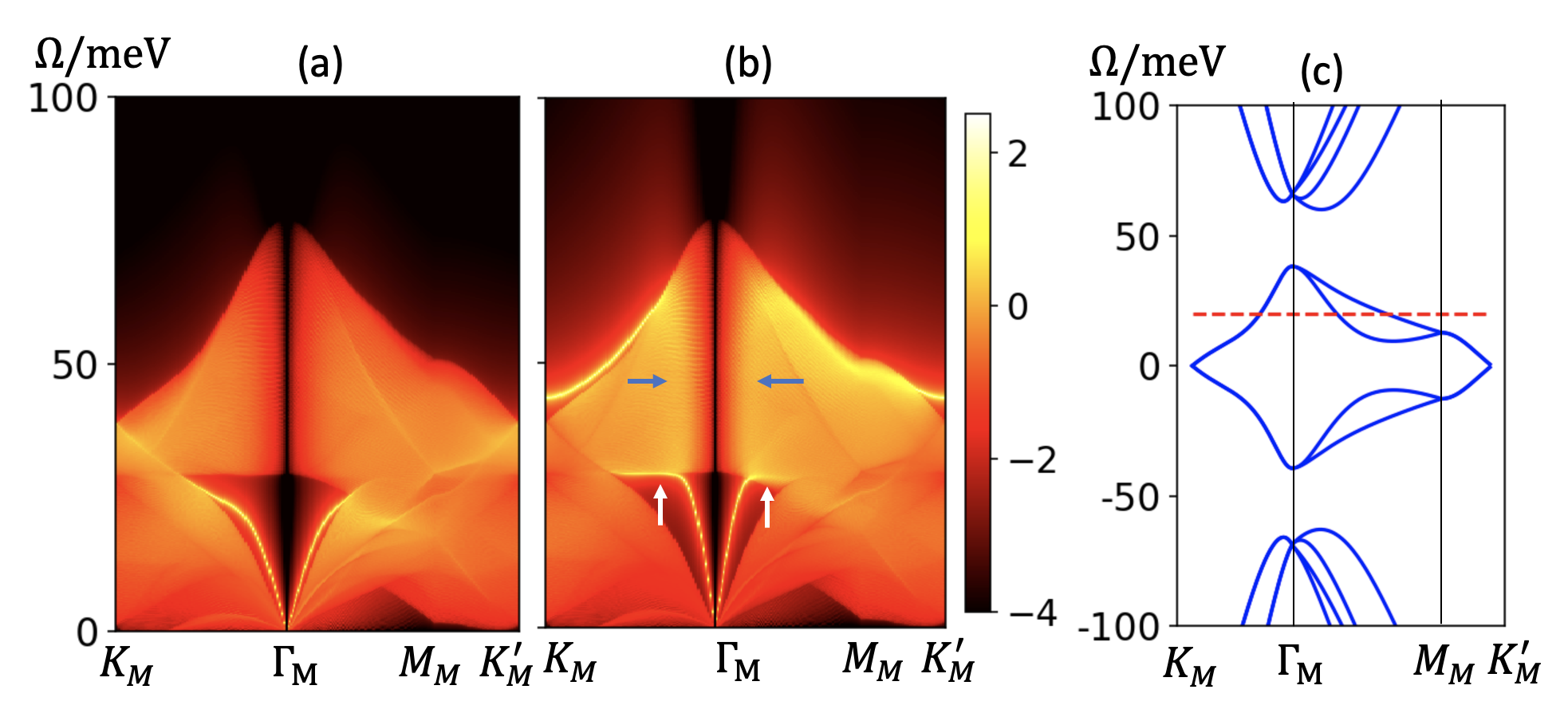}
\caption{\label{tbg_1.4} Numerical results for the dielectric function (Eq.~\ref{df}), $\tn{log}~\tn{Im}[{\epsilon^{-1}(q,\Omega)}_{\vec{G}=0}]$, for the model with interactions projected to flat-bands, for twisted bilayer graphene at $\theta=1.4$ for (a) realistic parameters, and (b) enhanced Coulomb interaction (see main text). The level-repulsion associated with the plasmon is marked by white arrows. (c) Band structure at $\theta=1.4$, with dashed line denoting the chemical potential $E_F=20~\tn{meV}$.}
\end{figure}

{\it Outlook.-} A number of experiments have reported measurements of the plasmon mode at room temperature using terrahertz near-field microscopy in graphene-based structures, ranging from micro-ribbon \cite{JuNat11}, single layer \cite{FrankZS23, FrankNat16, Basov12, Ju2}, to even twisted bilayer graphene away from magic-angle \cite{FrankNat21}. These experiments have focused primarily on the low-doping regime \cite{FrankZS23, FrankNat16, Basov12, Ju2}, or at high-energies ($\sim 200~\tn{meV}$) where the inter-band plasmon lies between the flat and remote bands  \cite{FrankNat21, BasovNano}, respectively. These relatively high temperature measurements also have inevitable thermal boardening ($\sim 25~\tn{meV}$), which makes it challenging to resolve a broadened plasmon that is repelled outside the continuum. However, with further technical advances in optical microscopy and identification of the ideal dielectric environment, future experiments will hopefully be able to observe an avoided level-repulsion and an undamped plasmon. 

On the theoretical front, a numerically exact investigation of the coupled dynamics of an intrinsic collective mode and the many-particle continuum in electronic models beyond an RPA-like expansion, used very commonly also in  other settings related to ours \cite{Maiti_2011}, is clearly desirable. A possible path would be to investigate correlated one-dimensional lattice models using time-dependent density matrix renormalization group techniques. Finally, whether a possible generalization of the same mechanism for avoiding (partially) a decay of the collective mode in non-Fermi liquid metals without long-lived quasiparticles \cite{me,Andrey3,zaanen} exists, remains an interesting open question.

{\it Acknowledgements.-} We thank K. Schalm and J. Zaanen for discussions, and especially A. Chubukov and C. Lewandowski for a number of useful suggestions based on a critical reading of an earlier version of this manuscript. XW thanks J.F. M\'endez-Valderrama for a number of useful suggestions related to the numerical simulations. DC acknowledges the hospitality of the Max Planck Institute for the Physics of Complex Systems during the final stages of this work. This work is supported in part by a CAREER grant from the NSF to DC (DMR-2237522) and by the Deutsche Forschungsgemeinschaft  under grant cluster of excellence ct.qmat (EXC 2147, project-id 390858490).  

\bibliographystyle{apsrev4-1_custom}
\bibliography{main.bib}

\clearpage
\renewcommand{\thefigure}{S\arabic{figure}}
\renewcommand{\figurename}{Supplemental Figure}
\setcounter{figure}{0}
\begin{widetext}
\begin{center}{\bf SUPPLEMENTARY INFORMATION for ``Interaction-mitigated Landau damping"}\end{center}

\title{SUPPLEMENTARY INFORMATION for ``Interaction-mitigated Landau damping"}

\author{Xuepeng Wang}
\affiliation{Department of Physics, Cornell University, Ithaca, New York 14853, USA.}
\author{Roderich Moessner}
\affiliation{Max-Planck-Institut f\"ur Physik komplexer Systeme, N\"othnitzer Stra\ss e 38, 01187 Dresden, Germany}
\author{Debanjan Chowdhury}
\affiliation{Department of Physics, Cornell University, Ithaca, New York 14853, USA.}

\maketitle

\section{Low-energy theory for the linearized Fermi surface coupled to gapped collective mode}\label{AppA}

In this section, we further elaborate on some of the technical details for the model introduced in Eq.~\ref{toymodel}. For simplicity, we set $\tilde{\rho}_s=0$  
in the following calculation. Let us start with Eq.~\ref{Dqo} and \ref{z_eq} for the $\Phi-$propagator and the associated quasiparticle residue: 
\begin{subequations}
\beq
D(\q,\Omega)&=&\frac{1/v_c^{2}}{\bigg[\widetilde{\Omega}^{2}-(\tilde{\Delta}^{2} - \widetilde{\lambda}) - \frac{\tilde{\lambda}\widetilde{\Omega}}{\sqrt{\widetilde{\Omega}^2-\q^2}}\bigg]},\label{Dqo_sup}\\
Z_\vp(\q) &=& \bigg[1+\tilde{\lambda} \frac{\q^2}{2\tilde{\Omega}_{\rm{peak}}(\tilde{\Omega}_{\rm{peak}}^2-\q^2)^{3/2}}\bigg]^{-1}.\label{z_eq_sup}
\eeq
\end{subequations}
In the regime $\tilde{\lambda}\gg 1$, we make the ansatz that the collective mode is expelled sufficiently far away from the upper edge of the ph continuum such that $\widetilde{\Omega}_{\rm{peak}}\gg |\q| \sim \tilde{\Delta}$, where $\Omega_{\rm{peak}}$ denotes the renormalized dispersion for $\Phi$ (obtained as the solution to $D^{-1}(\q,\Omega_{\rm{peak}})=0$). We then solve for the pole of Eq.~\ref{Dqo_sup} analytically by expanding $\Pi_c(\q,\Omega)$ in Eq.~\ref{pic} in powers of $q/\Omega$, leading to
\beq\label{gap_supA}
\widetilde{\Omega}^{2}-\widetilde{\Delta}^{2}-\tilde{\lambda}\frac{\q^2}{2\widetilde{\Omega}^2}=0.
\eeq
In the limit where $\widetilde{\Omega}_{\rm{peak}}\gg |\q| \sim \widetilde{\Delta}$ and $\tilde{\lambda}\gg1$, we can drop the $\widetilde{\Delta}^{2}$ in Eq.~\ref{gap_supA}, such that
\beq
\label{peak_sup}
\widetilde{\Omega}_{\tn{peak}}\simeq\bigg(\frac{\tilde{\lambda}\q^2}{2}\bigg)^{1/4}.
\eeq
To check for this behavior in the numerical data, for $\tilde\lambda=30$ we use a fit function $\tilde\Omega_{\tn
{peak}}\sim A q^\nu$, and find $\nu\approx0.53$. In the same asymptotic limit, this also immediately leads to $Z_\vp(\q)\rightarrow1/2$, for which we have obtained direct numerical evidence.

For decreasing $\tilde{\lambda}^2$, but in the regime where $\widetilde{\Omega}_{\tn{peak}}\gg q \sim \widetilde{\Delta}$, we can now include the perturbative corrections due to a small $\widetilde{\Delta}^{2}$ in Eq.~\ref{gap_supA}. At leading order, we obtain the correction to Eq.~\ref{peak_sup} as
\beq
\label{peak_sup_corr}
\widetilde{\Omega}_{\tn{peak}}\simeq\bigg(\frac{\tilde{\lambda}\q^2}{2}\bigg)^{1/4} + \frac{\tilde\Delta^2/2}{(\frac{\tilde{\lambda}\q^2}{2})^{1/4}}.
\eeq
Thus, the term $(\tilde\Omega^2_{\tn{peak}}-\q^2)$ in the denominator for $Z_\vp$ in Eq.~\ref{z_eq_sup} can be expressed in terms of
\beq\label{gap_sup}
\widetilde{\Omega}^2_{\tn{peak}}-|\q|^2\simeq\bigg(\frac{\tilde{\lambda}\q^2}{2}\bigg)^{1/2}+\Tilde\Delta^{2}-|\q|^2 = \bigg(\frac{\tilde{\lambda}\q^2}{2}\bigg)^{1/2}+\delta\omega(q)
\eeq
where $\delta\omega(q)=q_*^2-|\q|^2$ is subleading to the first term in the RHS of Eq.~\ref{gap_sup} ,and by definition $q_* = \Tilde\Delta$.
Note that the sign of $\delta\omega(q)$ in Eq.(\ref{gap_sup}) depends on whether $q<q_{\star}$ or $q>q_{\star}$, which gives rise to different branch when evaluating the expression in Eq.(\ref{z_eq_sup}). With the above simplifications, we obtain the expression for $Z_\vp$ in the main text, 
\beq\label{Z_sup_eq}
Z_\vp(\q) = \bigg[1 + \frac{(\frac{\tilde{\lambda}}{2}\q^2)^{3/4}}{[(\frac{\tilde{\lambda}}{2}\q^2)^{1/2} + \delta\omega(q) ~]^{3/2}} \bigg]^{-1}
=
\bigg[1 + \frac{\tilde{\lambda}^{3/4}}{[\tilde{\lambda}^{1/2} +(2/q^2)^{1/2} \delta\omega(q) ~]^{3/2}} \bigg]^{-1}.
\eeq

Let us now turn to the weak-coupling limit $\tilde{\lambda}\ll 1$. We discuss the asymptotic behavior of $Z_\vp$ in the limit $q \gg \widetilde\Delta$ and $q \ll \widetilde\Delta$, which corresponds to the regime marked by the blue and orange dashed line, respectively, in  Fig.~\ref{toy}b. For $q \gg \widetilde{\Delta}$ (blue dashed line), we assume that the splitting ($\Delta_g$) between the collective mode and the ph continuum is small, and expand $\widetilde{\Omega}_{\tn{peak}}=q+\Delta_g$, where $\Delta_g(\ll \widetilde{\Delta})$. Solving for the pole in Eq.~\ref{Dqo_sup},
\beq
\label{gap_sup_weak1}
(q+\Delta_g)^2-\widetilde{\Delta}^{2} + \tilde{\lambda} -\tilde{\lambda}\frac{q+\Delta_g}{\sqrt{(\Delta_g+q)^{2} -q^2}}=0, 
\eeq
which leads to (after dropping $\sim\Delta_g,~\widetilde{\Delta}$ terms), 
\beq\label{gap_sup_small}
\Delta_g \approx \frac{\tilde{\lambda}^2}{2q^{3}}.
\eeq
Note that Eq.~\ref{gap_sup_small} ($\Delta_g\sim\tilde\lambda^2\sim \lambda^4$) differs from the result from standard perturbation theory ($\sim\lambda^2$) between two discrete levels due to the divergent density of states $1/\sqrt{\Omega-q}$ near the onset of the continuum \cite{ruben18}. Approximating $\widetilde{\Omega}_{\rm{peak}}-q= \tilde{\lambda}^2/2q^{3}$ and $\tilde{\Omega}_{\rm{peak}}+q \approx 2q$, we obtain
\beq
Z_\vp^{-1} \approx \bigg[1+\frac{\q^{4}}{2\tilde{\lambda}^2}\bigg].
\eeq
On the other hand, when $q \ll \widetilde{\Delta}$ (see  dashed orange dashed line in Fig.~\ref{toy}b), we assume that $\widetilde{\Omega}_{\tn{peak}}\sim\widetilde{\Delta}$. Thus, we approximate $\widetilde{\Omega}\pm q \simeq \widetilde{\Delta}$, leading to
\beq
Z_\vp^{-1} \approx\bigg[1+\frac{\q^2}{2\widetilde{\Delta}^{4}}\tilde{\lambda}\bigg].
\eeq

Finally, the asymptotic expressions for $Z_\vp$ in the different regimes considered above can be summarized as,
\begin{subequations}
\beq
\label{zb_ana_sup}
Z_{\vp}(q<q_\star) &\approx& 
\begin{cases}\bigg[1+\frac{|\q|^2}{2\tilde{\Delta}^{4}}\tilde\lambda\bigg]^{-1} \sim 1 - \frac{\tilde\lambda|\q|^2}{2\tilde{\Delta}^{4}}, ~~\tilde\lambda \ll 1,
\\
\bigg(1+\frac{\tilde\lambda^{3/4}}{(\tilde\lambda^{1/2}+\delta q)^{3/2}}\bigg)^{-1}, ~~\tilde\lambda \gg 1,
\end{cases}\\
Z_{\vp}(q>q_\star) &\approx& 
\begin{cases}
\bigg[1+\frac{\q^{4}}{2\tilde\lambda^2}\bigg]^{-1}\sim\frac{2\tilde\lambda^2}{\q^{4}}, ~~\tilde\lambda \ll 1,\\
\bigg(1+\frac{\tilde\lambda^{3/4}}{(\tilde\lambda^{1/2}-\delta q)^{3/2}}\bigg)^{-1}, ~~\tilde\lambda \gg 1,
\end{cases}
\eeq
\end{subequations}
where we define $\delta q \equiv (2/q^2)^{1/2} |\delta \omega(q)|$ for simplicity.

\section{Random-flavor model in the Large$-N$ limit}
\label{App:random_flavor}
The exact Schwinger-Dyson equations for the boson and fermion Green's functions in the large$-N$ limit are given by
\begin{equation}
\begin{aligned}
\label{sde_yukawa}
G_c(\k,i\omega)=\frac{1}{i\omega_n-\epsilon_\k-\Sigma_{c}(\k,i\omega)},\\
D(\q,i\Omega)=\frac{1}{\Omega_n^2+\omega_\q^2+\Pi(\q,i\Omega)},
\end{aligned}    
\end{equation}
where the self-energies can be expressed graphically as
\begin{equation}\label{se_yukawa}
\begin{aligned}
\begin{tikzpicture}[x=0.75pt,y=0.75pt,yscale=-1,xscale=1]

\draw [color={rgb, 255:red, 0; green, 0; blue, 0 }  ,draw opacity=1 ]   (219.67,118.63) .. controls (199.09,118.87) and (222.09,118.87) .. (208.09,118.87) ;
\draw [color={rgb, 255:red, 0; green, 0; blue, 0 }  ,draw opacity=1 ]   (276.06,118.32) .. controls (264.67,91.06) and (229.67,92.06) .. (219.67,119.06) ;
\draw  [color={rgb, 255:red, 0; green, 0; blue, 0 }  ,draw opacity=1 ][fill={rgb, 255:red, 0; green, 0; blue, 0 }  ,fill opacity=1 ] (252.53,98.27) -- (246.95,102.23) -- (246.76,94.61) -- cycle ;
\draw [color={rgb, 255:red, 0; green, 0; blue, 0 }  ,draw opacity=1 ]   (276.06,118.32) .. controls (275.96,120.87) and (274.68,121.9) .. (272.22,121.41) .. controls (270.33,120.21) and (268.71,120.51) .. (267.34,122.32) .. controls (265.55,123.99) and (263.84,123.94) .. (262.21,122.18) .. controls (260.79,120.39) and (259.17,120.24) .. (257.36,121.73) .. controls (255.47,123.22) and (253.75,123.07) .. (252.21,121.26) .. controls (250.67,119.49) and (249.02,119.41) .. (247.27,121) .. controls (245.64,122.65) and (244.01,122.66) .. (242.38,121.02) .. controls (240.61,119.41) and (238.93,119.48) .. (237.34,121.23) .. controls (235.71,122.97) and (234.04,123.04) .. (232.34,121.44) .. controls (230.71,119.79) and (229.03,119.77) .. (227.3,121.38) .. controls (225.37,122.83) and (223.77,122.52) .. (222.5,120.46) -- (219.6,118.55) ;
\draw [color={rgb, 255:red, 0; green, 0; blue, 0 }  ,draw opacity=1 ]   (287.64,118.07) .. controls (267.06,118.32) and (290.06,118.32) .. (276.06,118.32) ;
\draw [color={rgb, 255:red, 0; green, 0; blue, 0 }  ,draw opacity=1 ] [dash pattern={on 0.84pt off 2.51pt}]  (276.06,118.32) .. controls (268.06,147.32) and (229.67,145.06) .. (219.67,119.06) ;
\draw  [color={rgb, 255:red, 0; green, 0; blue, 0 }  ,draw opacity=1 ][fill={rgb, 255:red, 0; green, 0; blue, 0 }  ,fill opacity=1 ] (284.64,118.07) -- (279.02,121.97) -- (278.91,114.35) -- cycle ;
\draw  [color={rgb, 255:red, 0; green, 0; blue, 0 }  ,draw opacity=1 ][fill={rgb, 255:red, 0; green, 0; blue, 0 }  ,fill opacity=1 ] (216.83,118.6) -- (211.21,122.49) -- (211.09,114.87) -- cycle ;
\draw    (342.67,113.56) .. controls (344.54,112.12) and (346.19,112.33) .. (347.63,114.2) .. controls (349.06,116.07) and (350.71,116.28) .. (352.58,114.84) .. controls (354.45,113.4) and (356.1,113.61) .. (357.54,115.48) -- (358.67,115.63) -- (358.67,115.63) ;
\draw    (411.85,115.32) .. controls (413.52,113.66) and (415.19,113.67) .. (416.85,115.34) .. controls (418.51,117.01) and (420.18,117.02) .. (421.85,115.36) -- (426,115.38) -- (426,115.38) ;
\draw [color={rgb, 255:red, 0; green, 0; blue, 0 }  ,draw opacity=1 ]   (411.85,115.32) .. controls (406,119.06) and (399.1,120.66) .. (387,120.06) .. controls (374.9,119.46) and (362.99,120.54) .. (358.67,115.63) ;
\draw [color={rgb, 255:red, 0; green, 0; blue, 0 }  ,draw opacity=1 ]   (411.85,115.32) .. controls (402.36,98.06) and (373.13,97.06) .. (358.67,115.63) ;
\draw  [color={rgb, 255:red, 0; green, 0; blue, 0 }  ,draw opacity=1 ][fill={rgb, 255:red, 0; green, 0; blue, 0 }  ,fill opacity=1 ] (390.53,102.27) -- (384.95,106.23) -- (384.76,98.61) -- cycle ;
\draw  [color={rgb, 255:red, 0; green, 0; blue, 0 }  ,draw opacity=1 ][fill={rgb, 255:red, 0; green, 0; blue, 0 }  ,fill opacity=1 ] (388.66,117.76) -- (388.44,124.59) -- (382.25,120.14) -- cycle ;
\draw [color={rgb, 255:red, 0; green, 0; blue, 0 }  ,draw opacity=1 ] [dash pattern={on 0.84pt off 2.51pt}]  (415.06,114.88) .. controls (407.06,143.88) and (368.67,141.63) .. (358.67,115.63) ;

\draw (167,109.4) node [anchor=north west][inner sep=0.75pt]  [font=\normalsize]  {$\Sigma _{c} =$};
\draw (307,109.4) node [anchor=north west][inner sep=0.75pt]  [font=\normalsize]  {$\Pi =$};
\end{tikzpicture}
\end{aligned}    
\end{equation}
The solid (curly) lines denote $c-$ ($\vp-$) Green's functions and the dashed line denotes the ``disorder'' contraction. 
We define this model on a square lattice, with the concomitant lattice dispersion for the fermions. For the bosonic dispersion, we choose $\omega_\q^2 = \rho_s \q^2 +\Delta^{\new{2}}$. We solve Eqs.~\ref{sde_yukawa} and \ref{se_yukawa} fully self-consistently along the Matsubara axis numerically, and ensure a convergence of the Green's functions up to $O(\tn{10}^{\tn{-4}})$. We discretize the momentum space with a mesh of size $101\times101$; the Matsubara frequency sums are
carried out with a discrete grid-size of $N_\omega = 2^9$. We perform analytic continuation numerically using the Pad\'e approximation. 

For the data shown in Fig.~\ref{toy}c and Fig.~\ref{toy}d, we used numerical (linear) interpolation along the momentum axis. The raw data is reproduced in Fig.~\ref{rf_zq}a and Fig.~\ref{rf_zq}b, for the same set of parameters as in the main text.

 \begin{figure}[h!]
\includegraphics[width=84mm,scale=1]{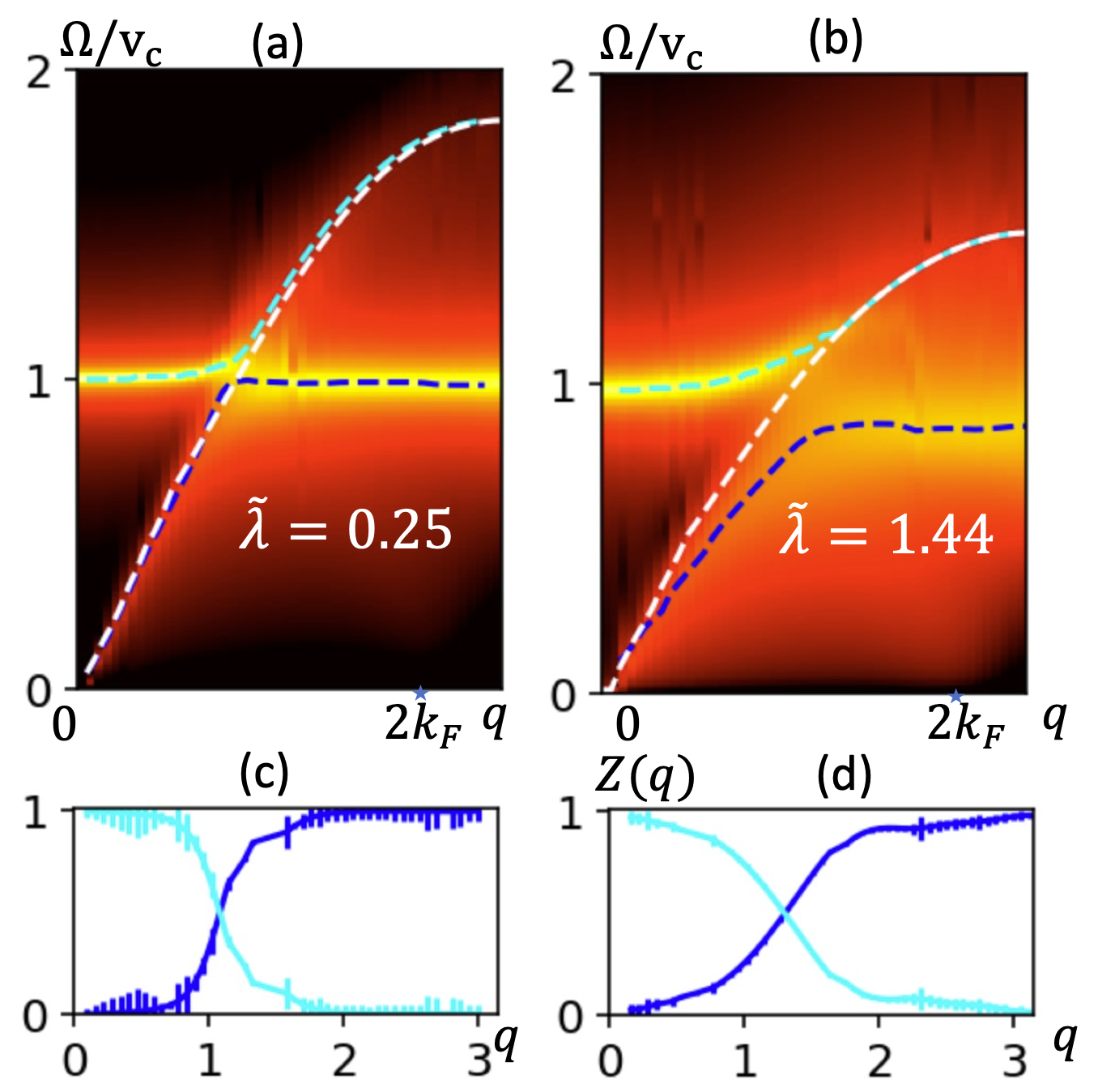}
\caption{\label{rf_zq} (a), (b) Unprocessed data without any linear interpolation for the response function, $\tn{log} [\tn{Im} D(\q,\Omega)]$, of the random flavor model. The parameter set is identical to that in Fig. \ref{toy}d, e, respectively. The cyan and blue dashed lines represent the branches used for fitting $\omega_{\tn{LD}}(q)$ and $\omega_{\tn{c}}(q)$ defined in Eq.\ref{double-lorentz}. The associated quasiparticle residues are obtained by fitting the numerical data to the functional form in Eq.\ref{double-lorentz}, smoothened by a gaussian filter with standard deviation $\sigma_q=0.2$. Error bars denote the error obtained from the fitting procedure.}
\end{figure}

To extract the quasiparticle residue, $Z_\vp$, as defined in Eq.~\ref{z_eq} numerically, we use the following functional form (Eq.~\ref{double-lorentz}) to fit the data for $D''(\q,\Omega)$ shown in Figs.~\ref{rf_zq}a, b.
\beq
\label{double-lorentz}
F''(\Omega) = A_{\tn{LD}}\frac{\sigma_{\tn{LD}}}{(\Omega-\omega_{\tn{LD}})^2+\sigma_{\tn{LD}}^2} + A_{\tn{c}}\frac{\sigma_{\tn{c}}}{(\Omega-\omega_{\tn{c}})^2+\sigma_{\tn{c}}^2}.
\eeq
Here the subscripts $\tn{LD}$ and $\tn{c}$ stand for Landau damping and collective mode, respectively. We fix $\omega_{\tn{c}}$ ($\omega_{\rm{LD}}$) to be along the dashed cyan (blue) line, respectively. The quasiparticle residues associated with these two branches, $A_{\tn{c}}$ and $A_{\tn{LD}}$, are plotted in Figs.~\ref{rf_zq}c and d. We note that our fitting procedure is only meant to capture the physics at an approximate level, especially since we do not include the effect of Landau damping explicitly in our functional form.

Finally, we argue that the multi-particle decaying processes shown in Fig.~\ref{toy}f  account for the decay of the collective mode even when it is expelled outside the ph continuum. Consider for simplicity the incoming boson with finite frequency and zero momentum, which scatters into a Landau damped boson with $|v_F^{*}\q|>\Omega$ as shown in Fig.~\ref{toy}f. The resulting self-energy can be written as,

\begin{equation}
\begin{aligned}
\label{multi_pi}
    \Pi_{\rm{multi}}(\q=0,i\Omega)&=\lambda^2\frac{N}{M}\int_{\omega_n,\k} \bigg(\frac{1}{i\omega_n-\epsilon_k}\bigg)^2\bigg[-i\omega_n (Z_c^{-1}-1)-i\alpha \omega_n^2 \tn{sgn}(\omega_n)\bigg]\frac{1}{i\omega_n+i\Omega_n-\epsilon_k}\\
    &=2\pi i \frac{\lambda^2}{v_F} \frac{N}{M} \int_{\omega_n} \frac{1}{\Omega_n^2}[\tn{sgn}(\omega_n+\Omega_n)-\tn{sgn}(\omega_n)][i\omega_n (Z_c^{-1}-1)+i\alpha \omega_n^2 \tn{sgn}(\omega_n)]\\
    &=\frac{2\pi\lambda^2}{v_F} \frac{N}{M} [(Z_c^{-1}-1)+\frac{2}{3}\alpha\Omega_n],
\end{aligned}
\end{equation}
where the terms in the square bracket in the first line of Eq.~\ref{multi_pi} arise from the fermionic self-energy in the FL phase; and $Z_c$ is the Fermi residue generated self-consistently from solving the saddle point equation Eq.~\ref{sde_yukawa}. Note that in our numerical computations we take $N=M$. For the sake of completeness, we reproduce the explicit computation of the FL self-energy after linearizing the dispersion near the Fermi surface:
\beq\label{FL_se}
\bal
\Sigma_{\tn{FL}}(\q,i\Omega)&=\lambda^2\int_{\omega,k_x,k_y}\frac{1}{i(\omega_n+\Omega_n)-v_F(k_x+q_x)-\kappa(k_y+q_y)^2}\frac{k_y}{\tilde\lambda |\omega| - \Delta^{\new{2}} k_y}\\&
=2\pi i~\frac{\lambda^2}{v_F}\int_{\omega,k_y}\frac{\tn{sgn}(\omega+\Omega)~k_y}{\tilde\lambda |\omega| - \Delta^{\new{2}} k_y} =-2\pi i\frac{\lambda^2}{\Delta^{\new{2}} v_F}\int_{\omega}[k_F+\frac{\tilde\lambda}{\Delta^{\new{2}}}|\omega|]\tn{sgn}(\omega+\Omega)\\&
=\bigg[-i\Omega_n (Z_c^{-1}-1)-i\alpha \Omega_n^2 \tn{sgn}(\Omega_n)\bigg],
\eal
\eeq
where $(Z_c^{-1}-1)\sim\lambda^2/v_F\Delta^{\new{2}}$, and $\alpha\sim\lambda^4/v_F^3 \Delta^{\new{4}}$. We ignore the logarithmic correction in the second line of Eq.~\ref{FL_se}. Thus, after analytical continuation, we obtain $\Pi_{\tn{multi}}(q,\Omega)=2\pi\lambda^2 [(Z_c^{-1}-1)+i\alpha\Omega]/v_F$, which gives finite lifetime $i\lambda^2 \alpha \Omega/v_F \sim O(\lambda^6)\sim O(\tilde\lambda^3)$. 

\section{Particle-hole Continuum in Graphene}\label{App:graphene}
In this section, we  provide additional details related to analytical results for the particle-hole continuum of graphene \cite{rmp_sarma,prb_sarma}. 
The form-factors associated with the graphene pseudospinor wavefunctions, $\Psi^s_\k$, are defined as $F^{s,s'}(\q,\k)=|\langle \Psi^{s}_{\k+\q} | \Psi^{s'}_{\k} \rangle|^2$, where $s$ and $s'$ represent the band index. In the long-wavelength limit, the intraband and interband form-factors are given by \cite{pnas}, 
\begin{equation} 
\begin{aligned}
&F_{\tn{\tn{intra}}} (\q,\k) \equiv F^{s,s}_{\k+\q,\k} \simeq   1 \\
&F_{\tn{\tn{inter}}} (\q,\k) \equiv F^{s,\bar{s}}_{\k+\q,\k} \simeq \frac{\q^2}{\k^2} \sin^2\theta_\q,
\end{aligned}
\end{equation}
where $\theta_\q$ is the angle between $\k$ and $\q$. The particle-hole bubble is given by,
\beq
\Pi^{s,s'}(\q,\Omega)=\int d^2\k ~ F^{s,s'}_{\k+\q,\k} \frac{n_F(\epsilon^{s}_{\k+\q})-n_F(\epsilon^{s'}_{\k})}{\epsilon^{s}_{\k+\q}-\epsilon^{s'}_{\k}-\Omega}.
\eeq

The intraband continuum defined as $\Pi_{\tn{intra}}(\q,\Omega) \equiv \Pi^{s,s}(\q,\Omega)$ is given by 
\begin{equation}\label{asym_intra}
\begin{aligned}
\Pi_{\tn{intra}}(\q,\Omega)&=\int_{\k}\frac{F_{\tn{intra}}(\k)~[f_{-}(\k)-f_{-}(\k+\q)]}{\Omega-(\epsilon_{\k+\q}-\epsilon_{\k})}\\
&\simeq-\int_{\k}\frac{F_{\tn{intra}}(\k)~\delta(\epsilon_\k)v_F |\q| \cos \theta}{\Omega-v_F |\q| \cos \theta}\\
&=\frac{1}{v_F}\bigg[1-\frac{\Omega}{\sqrt{\Omega^2-(v_F |\q|)^2}}\bigg],
\end{aligned}
\end{equation}
which is the standard result for metallic systems in the low-energy limit (i.e. holds more generally for actions like Eq.~(\ref{action})).

Without loss of generality, we assume that the system is hole doped so that $n_F(\epsilon^+_{\k+\q})=0$ at $T=0$. We note that the results for $\Pi^{s,s'}(\q,\Omega)$ are the same if we switch from hole doping to electron doping with the same carrier density because the single-particle band structure, $\epsilon_\k^s = s v_F |\k|$, is symmetric. The interband continuum defined as $\Pi_{\tn{inter}}(\q,\Omega) \equiv \Pi^{s,\bar{s}}(\q,\Omega)$ is given by 
\begin{equation}\label{asym_inter}
\begin{aligned}
\Pi_{\tn{inter}}(\q,\Omega)&=\sum_{s}\int_{\k}\frac{F^{s,\bar{s}}_{\k+\q,\k}(\k)[n_F(\epsilon^{s}_{\k+\q})-n_F(\epsilon^{\bar{s}}_{\k})]}{(\epsilon^{s}_{\k+\q}-\epsilon^{\bar{s}}_{\k})-\Omega}\\
&=-\sum_{s}\int_{\k}n_F(\epsilon^{s}_{\k})F^{\bar{s},s}_{\k,\k+\q}\bigg[\frac{1}{\epsilon^{\bar{s}}_{\k+\q}-\epsilon^{s}_{\k}-\Omega} + \frac{1}{\epsilon^{\bar{s}}_{\k+\q}-\epsilon^{s}_{\k}+\Omega}\bigg]\\
&\simeq-\int_{\k}\bigg[\frac{n_F(\epsilon^{-}_{\k})}{2v_F |\k|-\Omega+v_F |\q| \cos \theta}+\frac{n_F(\epsilon^{-}_{\k})}{2v_F |\k|+\Omega+v_F |\q| \cos \theta}\bigg]~ \frac{|\q|^2}{|\k|^2}\sin^2 \theta_\q\\
&=\frac{\pi}{v_F^2}\int_{k_F}^{\Lambda} kdk~\bigg[\frac{\sqrt{(\Omega-2v_F k)^2-(v_F q)^2}}{k^2} + \frac{\sqrt{(\Omega+2v_F k)^2-(v_F q)^2}}{k^2} - \new{\frac{4v_F}{k}}\bigg]\\
&\sim \frac{\pi}{v_F^2}\bigg[\sqrt{(\Omega-2E_F)^2-(v_F q)^2}+\sqrt{(\Omega+2E_F)^2-(v_F q)^2}-4E_F\bigg]+\Pi_{\rm{reg}}(\q,\Omega)\\
&\sim \frac{\pi}{v_F^2}\bigg[\sqrt{(\Omega-2E_F)^2-(v_F q)^2}+(\Omega-2E_F)\bigg]+\Pi_{\rm{reg}}(\q,\Omega),
\end{aligned}
\end{equation}
where $\Pi_{\tn{reg}}(\q,\Omega)=-2\sqrt{\Omega^2-(v_F q)^2}\bigg[\tn{arctanh}\bigg(\frac{\sqrt{(\Omega+2E_F)^2-(v_F q)^2}+2E_F}{\sqrt{\Omega^2-(v_F q)^2}}\bigg) + \tn{arctanh}\bigg(\frac{\sqrt{(-\Omega+2E_F)^2-(v_F q)^2}+2E_F}{\sqrt{\Omega^2-(v_F q)^2}}\bigg)\bigg]+\Omega\tn{log}\bigg[\frac{\sqrt{(\Omega+2E_F)^2-(v_F q)^2}+\Omega+2E_F}{\sqrt{(-\Omega+2E_F)^2-(v_F q)^2}-\Omega+2E_F}\bigg]$ is non-singular in $(\Omega-2E_F-v_F q)$. 
To illustrate the qualitative agreement between the results of the full numerical computation and the analytical asymptotic analysis with regards to the plasmon level-repulsion, we plot $\tn{Im} [1/\epsilon(\q,\Omega)]$ in Fig.~\ref{graphene_sup}, defined as
\beq
\frac{1}{\epsilon(\q,\Omega)}=\frac{1}{1-\frac{\beta_0 |\q|}{\tilde\Omega^2}-\frac{\tilde{e}^2}{|\q|}\Pi_{\tn{inter}}(\q,\Omega)},
\eeq
where we use the asymptotic expression for $\Pi_{\tn{inter}}(\q,\Omega)$ obtained in Eq.~\ref{asym_inter}. The plasmon dispersion without coupling to the inter-band continuum is shown in Fig. ~\ref{graphene_sup}b, by setting $\tilde e^2 =0 $. In contrast, after coupling to the inter-band continuum (for finite $\tilde e^2$), the plasmon mode is pushed towards lower energy, highlighted by the white arrow in Fig.~\ref{graphene_sup}.

\begin{figure}[h!]
\includegraphics[width=160mm,scale=1]{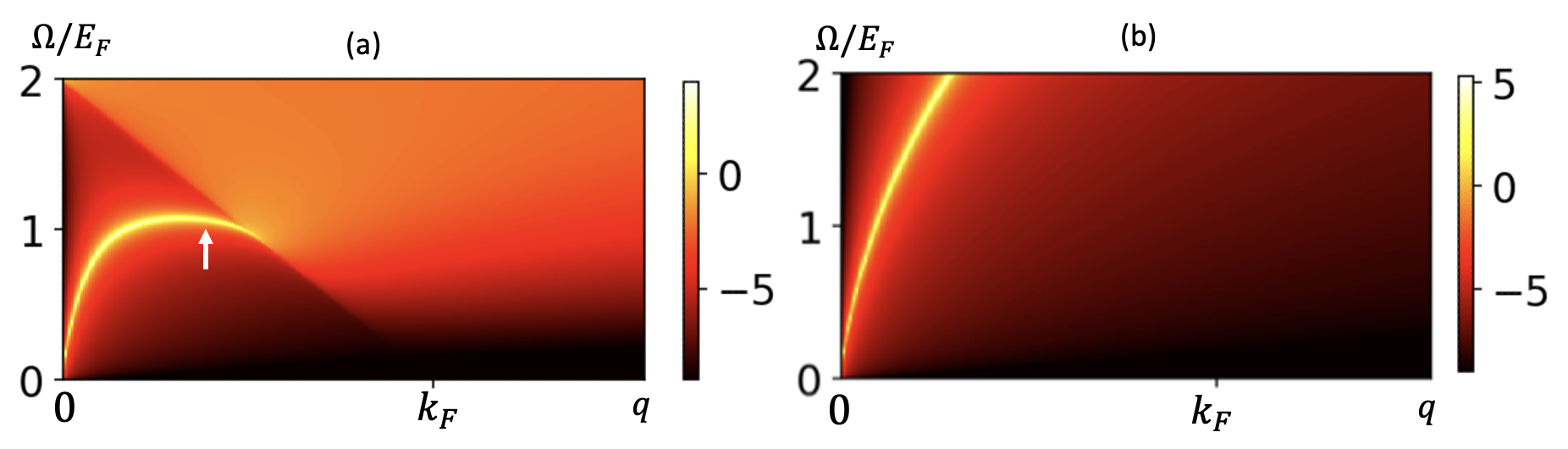}
\caption{\label{graphene_sup} Plot of $\tn{log}[ \tn{Im} \chi (\q,\Omega)]$ (see Eq.(\ref{graphene_asym})) based on the analytical forms in Eqs.~\ref{asym_intra}-\ref{asym_inter} with (a) $\tilde{e}^2=8$, and (b) $\tilde{e}^2=0$. The white arrow highlights level-repulsion of the plasmon mode, analogous to the full numerical computation in Fig.~\ref{graphene_main}.}
\end{figure}

Our discussion of the phenomenology of avoided plasmon decay in (moir\'e) graphene has mostly focused on the interband continuum. Here we make some brief remarks regarding the coupling between the  plasmon and the intraband continuum. For simplicity, consider the effect of introducing the three-dimensional Coulomb interaction, $V(\q)\sim1/q^2$, which has the advantage of gapping out the plasmon. Moreover, we can ignore the contribution from $\Pi_{\tn{inter}}(q,\Omega)$. The dielectric function is given by,
\beq\label{eps_intra}
\frac{1}{\epsilon(\q,\Omega)}=\frac{|\q|^2}{|\q|^2+ e^2 \Pi_{\tn{intra}}(\q,\Omega)}.
\eeq
By solving for $\epsilon(\q,\Omega)=0$, the dispersion of intrinsic plasmon mode $\Omega_{\tn{peak}}(\q)$ is obtained as
\beq
\Omega_{\tn{peak}}(\q)=\frac{(v_F \q^2 + e^2)\sqrt{v_F}}{\sqrt{v_F \q^2 + 2e^2}} = \bigg[ \frac{v_F\q^2 + e^2}{\sqrt{(v_F\q^2 + e^2)^2 - e^4}} \bigg]v_F |\q| > v_F |\q|.
\eeq
By comparing $\Omega_{\tn{peak}}(|\q|)$ with the onset of ph continuum $v_F |\q|$, we find that the collective mode is {\it completely} repelled outside the continuum, which agrees with previous studies \cite{sarma1, sarma2}.

\section{Density response in twisted bilayer graphene}\label{App:tbg}

We start by briefly reviewing the Bistritzer-Macdonald (BM) continuum model. Consider two layers of the graphene Hamiltonian at low energy in momentum space, where each layer consists of two Dirac cones at $\K_{\eta}^{l}$ with $\eta=\pm$ denoting the valley  and $l=1,2$ the layer index. A relative twist for these two layers in real space by an angle $\theta$ results in the momentum space rotation by the same angle. The two Dirac points in different layers and the same valley are separated by $\frac{4\pi}{3L_M}$, with the moir\'e reciprocal lattice vectors given by
\begin{equation}
\begin{aligned}
    \vec{G^{M}_1} = \frac{2\pi}{\sqrt{3} L_M} 
\begin{pmatrix}
1  \\
\sqrt{3}  
\end{pmatrix},
~~~
\vec{G^{M}_2} = \frac{4\pi}{\sqrt{3} L_M} 
\begin{pmatrix}
1  \\
0  
\end{pmatrix},
\end{aligned}
\end{equation}
where $L_M\simeq a/\theta$ is the lattice constant for the moir\'e unit cell and $a$ is the lattice constant for the original graphene layer.
The low energy Dirac Hamiltonian $H^{\eta}_{l}(\theta)$ is given by
\begin{equation}
\begin{aligned}
H^{\eta}_{l}(\theta)=-v_F[(\k - \K^{l}_{\eta})\cdot R^\dagger(\pm \theta/2)(\eta \sigma_x, \sigma_y) R(\pm \theta/2)],
\end{aligned}
\end{equation}
where $R(\pm \theta/2)$ are the rotation matrix $R_{\theta}=e^{i \frac{\theta}{2} \sigma_z}$. For each valley, after including the inter-layer hopping, the Hamiltonian is
\begin{equation}
\begin{aligned}
\label{ham_bm}
    H^{\eta}=\begin{pmatrix}
H_1(\theta/2) & U^{\dagger} \\
U & H_2(-\theta/2) 
\end{pmatrix},
\end{aligned}
\end{equation}
where the inter-layer hopping in real space is given by
\begin{equation}
\begin{aligned}
    U=
\begin{pmatrix}
w_0 & w_1 \\
w_1 & w_0
\end{pmatrix}
+
\begin{pmatrix}
w_0 & w_1 e^{-i 2\pi/3} \\
w_1 e^{i 2\pi/3} & w_0
\end{pmatrix} e^{i\eta \vec{G^{M}_1}\cdot r}
+
\begin{pmatrix}
w_0 & w_1 e^{i 2\pi/3} \\
w_1 e^{-i 2\pi/3} & w_0
\end{pmatrix}
e^{i\eta (\vec{G^{M}_1}+\vec{G^{M}_2})\cdot r}.
\end{aligned}
\end{equation}
For our numerical computations, we pick $w_0=0.0797~\tn{eV}$, $w_1=0.0975~\tn{eV}$ and $v_F/a=2.1354~ \tn{eV}$.

To calculate the RPA response function, we first diagonalize the Hamiltonian in Eq.~\ref{ham_bm} in the moir\'e Brillioun zone (mBZ). The creation operator in the band basis, $f^{\dagger}$, can be expressed in terms of the wave function $u^{n}_{\Q, \eta, l}(\k)$ as,
\be
\bal
f^{\dagger}_{\k,n,\eta}=\sum_{\Q, \alpha} u^{n}_{\Q, \eta, \alpha}(\k) c^{\dagger}_{\k,\Q,\eta,\alpha},
\eal
\ee
where $c^{\dagger}$ is the original creation operator. Here $n$ labels the band index, $\Q$ denotes the Dirac point honeycomb lattice, $\eta$ denotes valley, and $\alpha$ denotes sublattice index, respectively.

The projected Coulomb interaction is given by \cite{tbg3}
\beq
\bal
H_I=\sum_{\q\in \tn{mBZ}, \vec{G}} \sum_{\eta,\tilde{\eta}} \sum_{n,n',m,m'} V(\q+\vec{G}) ~M_{n,n'}^{\eta}(\k,\q+\vec{G}) M_{m,m'}^{*~\tilde{\eta}}(\tilde{\k},\q+\vec{G}) ~f^{\dagger}_{\k+\q,n,\eta} f_{\k,n',\eta} ~f^{\dagger}_{\tilde{\k}-\q,m',\tilde{\eta}} f_{\tilde{\k},m,\tilde{\eta}},
\eal
\eeq
where $\vec{G}$ is summed over the moir\'e reciprocal lattice, and the form factor is defined as $M_{n,n'}^{\eta}(\k,\q+\vec{G})=\sum_{l,\Q} u^{*}_{\Q-\vec{G},l,n}(\k) u_{\Q, l, n'}(\k)$. The projection onto the flat-band is carried out by restricting $n,n',m,m'=0,1$.

We define the particle-hole bubble diagram as
\beq
\Pi^{\eta}_{\vec{G},\vec{G'}} (\q,\Omega) = \sum_{\k,m,n,\eta} M^{* ~\eta}_{m,n} (\k, \q+\vec{G}) M^{\eta}_{m,n} (\k, \q+\vec{G'})
\frac{n_F(\epsilon_{\k+\q}^m) - n_F(\epsilon_{\k}^n)}{\epsilon_{\k+\q}^m - \epsilon_{\k}^n - \Omega},
\eeq
where $\epsilon_{\k}^n$ denotes the dispersion of the $n-$th band. The associated dielectric function is given by \cite{cea_pnas}
\beq
\label{trace_pi}
\epsilon^{-1}(\q,\Omega)_{\vec{G},\vec{G'}} = \bigg[\mathbf{1}_{\vec{\tilde G},\vec{\tilde G'}} - \sum_{\eta}V(\q+\vec{\tilde G}) \Pi^{\eta}_{\vec{\tilde G},\vec{\tilde G'}}(\q,\Omega)\bigg]^{-1}_{\vec{G},\vec{G'}},
\eeq
where the $[...]^{-1}$ denotes matrix inversion in the $\vec{\tilde G},~\vec{\tilde G'}$ space. Numerically, we enforce a cutoff for $\vec{G}$ to be within (including) the third nearest neighbour moir\'e unit cell, as the form factor decays exponentially with increasing number of umklapp wavevectors \cite{tbg3}; moreover, we only focus on the diagonal piece of $\Pi^{\eta}_{\vec{G},\vec{G'}}(\q,\Omega)$ (with $\vec{G}=\vec{G'}$). We note that the local field effect \cite{local1, local2, fahimniya2020dipoleactive, cyprian_sci_adv} is not included in our numerics, where we have ignored the off-diagonal piece in Eq.~\ref{trace_pi}. However, for $\q$ near $\Gamma$-point in the moir\'e Brillouin zone, where the level-repulsion is predicted, it is expected that the local field effect does not change the result qualitatively \cite{pnas, cyprian_sci_adv}. 

For the sake of completeness, we also compute numerically the contribution to ph continuum from the remote bands, and the off-diagonal pieces of $\Pi^{\eta}_{\vec{G},\vec{G'}}(\q,\Omega)$ for $\vec{G}$ and $\vec{G'}$ to be within (including) the nearest neighbour moir\'e unit cell. The corresponding results for $\tn{Im}~[\epsilon^{-1}(\q,\Omega)]_{\vec{G}=\vec{G'}=0}$ is shown in Fig.~\ref{local_field}.

\begin{figure}[h!]
\includegraphics[width=120mm,scale=1]{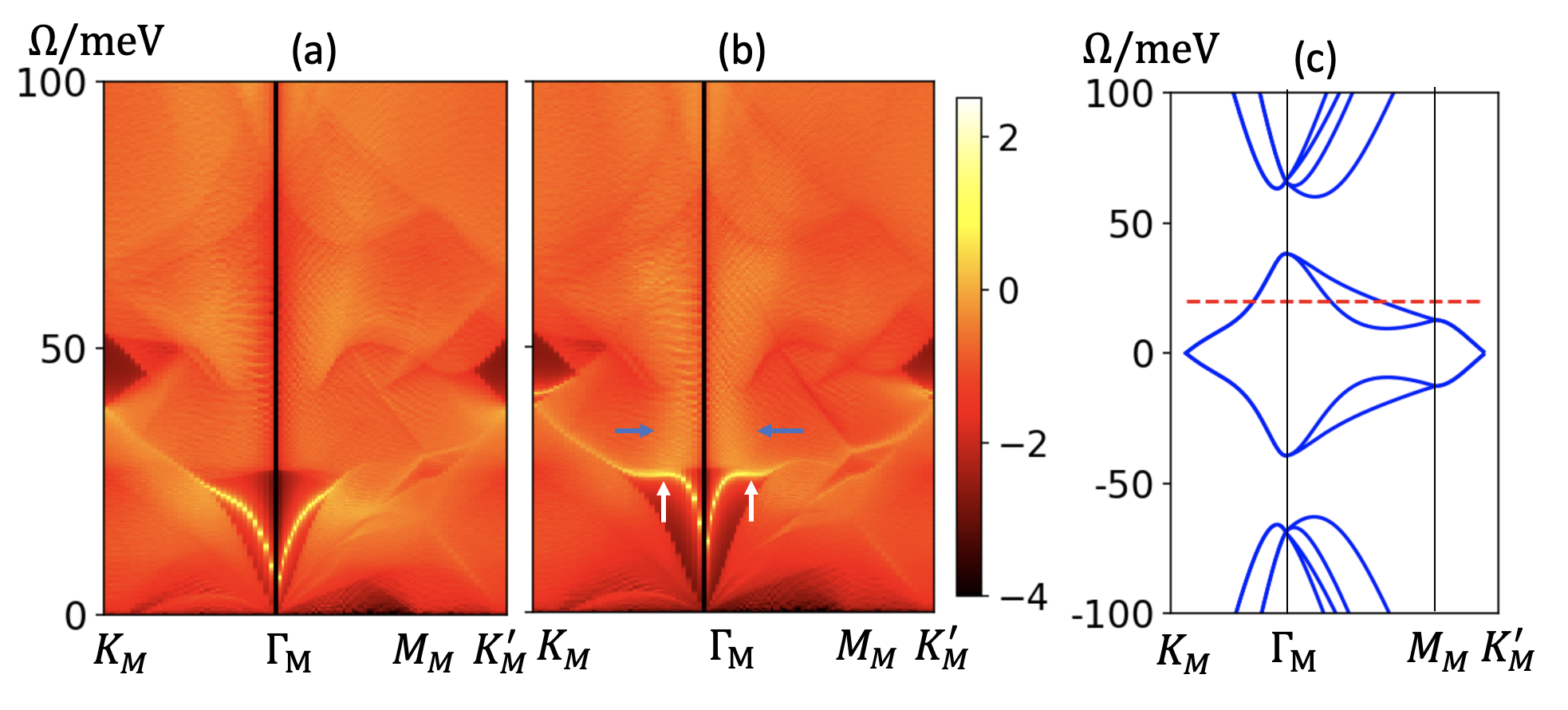}
\caption{\label{local_field} Numerical results for the dielectric function (Eq.~\ref{trace_pi}), $\tn{log}~\tn{Im}[{\epsilon^{-1}(q,\Omega)_{\vec{G}=\vec{G'}=0}}]$ with the inclusion of the remote bands and local field effect. The parameter set is identical to that in Fig. \ref{tbg_1.4} (a)-(c), respectively.}
\end{figure}

\end{widetext}

\end{document}